# Phase transition in crowd synchrony of delay-coupled multilayer laser networks

Elad Cohen[1], Michael Rosenbluh[1] and Ido Kanter[2,3*]

[1]Department of Physics, The Jack and Pearl Resnick Institute for Advanced Technology, Bar-Ilan University, Ramat-Gan 52900, Israel
[2]Department of Physics, Bar-Ilan University, Ramat-Gan 52900, Israel
[3]Gonda Interdisciplinary Brain Research Center, and the Goodman Faculty of Life Sciences, Bar-Ilan University, Ramat-Gan 52900, Israel.

[*]*ido.kanter@biu.ac.il*

An analogy between crowd synchrony and multi-layer neural network architectures is proposed. It indicates that many non-identical dynamical elements (oscillators) communicating indirectly via a few mediators (hubs) can synchronize when the number of delayed couplings to the hubs or the strength of the couplings is large enough. This phenomenon is modeled using a system of semiconductor lasers optically delay-coupled in either a fully connected or a diluted manner to a fixed number of non-identical central hub lasers. A universal phase transition to crowd synchrony with hysteresis is observed, where the time to achieve synchronization diverges near the critical coupling independent of the number of hubs.

Synchronization of the dynamics of non-identical oscillators interacting with each other through a common medium is observed in many phenomena in biology [1, 2] optics [3] and many other complex networks. This phenomenon attracted international notice in the year 2000, when the London Millennium Bridge caused great alarm due to large oscillations [4, 5] as pedestrians walking at initially random pace over the bridge were weakly coupled by the bridge's lateral oscillations. When the number of people walking on the bridge reached a critical value, synchronization between the bridge oscillations and the pedestrian's walking pace and phase emerged. In a recent pioneering paper [6], this phenomenon of crowd synchrony was investigated and generalized using a group of non-identical semiconductor lasers interacting via delay-coupling with a central hub laser. The solitary hub laser, operating well below lasing threshold, is a "passive" coupler, similar to the bridge in the absence of walking pedestrians. Similar quorum sensing phenomena have been observed in yeast cells [7] and in a chemical oscillator system [8], with synchronization depending on the density of the system.

There are two critical questions which need to be addressed. The first is whether the emergence of crowd synchrony is a phase transition when the number and strength of the couplings is increased and if so whether the transition is continuous or discontinuous. The second question examines the generality of network architectures for which crowd synchrony emerges, in particular, whether crowd synchrony can emerge in scenarios where elements interact indirectly via several hubs. Furthermore, can synchronization emerge in a partially connected system where the symmetry among the elements is broken, e.g. elements interact with different subsets and numbers of non-identical hubs. These questions are addressed in this Letter using tools and methods developed during the last two decades to analyze the capabilities of multilayer neural networks architectures [9, 10].

Figure 1a shows a two-layer architecture consisting of *M* non-identical elements operating at independent frequencies connected via *P* identical hubs, known in the realm of neural networks as hidden units (HUs). The dilution of the architecture is defined by a set of vanishing couplings, $\sigma_{ik}=0$, between element *i* and the k*th* HU, whereas all existing couplings have equal strength, $\sigma$. The dilution breaks the symmetry among the *M* elements since they are connected and influenced by different numbers and subsets of HUs. The limiting case of *P*=1 is the star architecture discussed in [6] where only a fully connected system is possible. The star architecture is known as the perceptron [9, 11] in the realm of neural network whose hidden unit feeds back to the input level rather than to an output unit. Figure 1b presents the simplest case of two non-identical HUs with dilution and only two types of coupling strengths. Expanding the bridge analogy to the multilayer architectures, one can imagine a group of pedestrians simultaneously walking on several identical bridges (Fig. 1a) or non-identical bridges (Fig. 1b).

For the two layered structure shown in Fig. 1 and following reference [6] the equations describing the slow envelope of the complex electric field of the i*th* element of the lower layer $E_i$, the k*th* element of upper layer $E_k$ and the carrier density $N_{i,k}$ of the lasers in the upper and lower layers are:

$$\dot{E}_i = iw_i E_i + \gamma(1 + i\alpha)(G_i - 1)E_i + \sum_{k=1}^{P} \sigma_{ik} E_k(t - \tau)e^{-iw_0\tau} + \sqrt{D}\zeta_i(t) \qquad (1)$$

$$\dot{E}_k = iw_k E_H + \gamma(1 + i\alpha) \cdot (G_k - 1)E_k + \sum_{i=1}^{M} \sigma_{ik} E_i(t - \tau)e^{-iw_0\tau} + \sqrt{D}\zeta_k(t) \qquad (2)$$

$$\dot{N}_{i,k} = \gamma_e (p_{i,k} - N_{i,k} - G_{i,k}|E_{i,k}|)^2 \qquad (3)$$

where $G_{j,k}=N_{j,k}/(1+\varepsilon|E_{j,k}|^2)$. The coupling delay time between the lasers is $\tau$, the field and carrier decay rates are $\gamma$ and $\gamma_e$, respectively, $w_0$ is the optical frequency and $w_k$ and $w_i$ are the detuning of HU lasers and lower layer lasers from $w_0$ $\alpha$ is the linewidth enhancement factor and $\varepsilon$ is the gain saturation. The uncorrelated complex Gaussian noise is represented by $\zeta_i$ and $\zeta_k$ with strength $D$. The normalized injection current of the lasers is $p_{i,k}$ where we assume that each laser has a normalized threshold current $p_{th}=1$. The chosen parameter values (typical semiconductor laser): $\gamma=300$ ns$^{-1}$, $\gamma_e=1$ ns$^{-1}$, $\alpha=3$, $D=10^{-5}$ ns$^{-1}$, $\omega_o=2\pi c/\lambda$ ($\lambda=654$ nm). We assume a uniform coupling delay of $\tau=5$ ns and the rate equations are calculated every $10^{-13}$ s using the Euler method. All bottom layer lasers are characterized by the same parameters with the only difference being the frequency detuning ($\omega_\iota$ is chosen from a Gaussian distribution with zero mean and standard deviation of $20\pi$ rad/ns) and random initial conditions. There are two cases for the lasers in the top layer; initially we set all frequency shifts to zero ($\omega_\kappa=0$) and later (Fig. 1b) we allow for a frequency difference. In the latter case without loss of generality the frequency shift of the first HU laser is set to be zero, $\omega_1=0$, while $\omega_2=a$. The HU lasers as well as the lasers in the lower layer are all assumed to have pump currents well below $p_{th}$, and in the simulations we investigate a range of injection currents for the lower/upper layers. The results presented here are for $p_i/p_k=0.7/0.4$, but we find similar results for other values as long as all lasers are operated below their solitary lasing threshold [6].

To measure synchronization in the lower layer it was suggested [6] to use an order parameter, $Q$, equal to the normalized time-averaged coherent intensity summed over all lasers, $Q=<I>/M$, where $<>$ denotes the average over a time window T. $Q$ is a constant in the absence of synchronization and grows with $M$ as crowd synchrony emerges. The critical crowd, $M_c$, was defined where the order parameter changes abruptly. Though this provides a convenient measure for synchronization for simple architectures, for more general architectures, such as those containing multiple HUs, this criterion is less appropriate, since cluster synchronization of a subset of the lasers might emerge. In addition, since this criterion looks at the change of $Q$ ($dQ/dM$) one cannot determine if a single simulation with a given $M$ is synchronized or not. Finally, since $M$ is an integer one cannot investigate the continuous mathematical nature of the transition to synchronization.

We suggest measuring the Pearson correlation, $\rho$, of the output intensity amongst all pairs of lower layer lasers over a time window of T:

$$\rho(i,j) = \frac{\sum_{t=0}^{T}(I_i(t) - \overline{I_i})(I_j(t) - \overline{I_j})}{\sqrt{\sum_{t=0}^{T}(I_i(t) - \overline{I_i})^2 \sum_{t=0}^{T}(I_j(t) - \overline{I_j})^2}}$$

Where $I_i(t)=|E_i(t)|^2$ is the *ith* laser's intensity at time $t$ and $\overline{I_i}=\Sigma^T_{t=0}I_i(t)/T$ is the average of $I_i(t)$ over $T$.

Synchronization of the lasers in the entire lower layer can be deduced from the histogram of $\rho(i,j)$ using one of the following threshold criteria: the mean value of all the correlations in the histogram or the minimal correlation value in the histogram (which is dominated by the correlation between laser pairs on opposite extremes of the frequency distribution). Using any of these criteria in our simulations yields similar qualitative results. To minimize computational complexity, our selected calculation criteria for synchronization was that the minimal value of correlations in the histogram be above a threshold value, $\rho=0.7$. Once synchronized, all lower level lasers receive roughly the same coupling amplitude, however, lasers on the extremes of the frequency distribution are less effectively coupled and as a result, their lasing intensities are lower. The time dependent intensity fluctuations of all lasers, however, are almost identical reflected by $\rho$ approaching unity, e.g. Fig. 2c.

Figure 2 shows the correlation matrix $\rho(i,j)$ for the fully connected architecture of Fig. 1a with $M=20$ and $P=3$. The continuous tuning parameter, as opposed to $M$, is the coupling strength, $\sigma$. The simulations indicate that this multilayer architecture undergoes a sharp transition to crowd synchrony. Cluster synchronizations are observed on the path to global synchronization, where lasers which possess closer frequencies first synchronize with each other. In order to identify the critical coupling strength, $\sigma_c$, we turn to a technique developed to identify the maximal capacity, (maximal number of input/output relations that can be embedded in a network), of multilayer neural networks based on the divergence of the learning time as criticality is approached [12, 13]. Similarly, the time needed to achieve crowd synchrony for a given architecture diverges as the coupling strength approaches $\sigma_c$ from above. In Figure 3a we show this time as a function of the coupling strength for a single HU and for multiple hubs with 3 HUs. Synchronization time follows a power law with an exponent ~-0.8, independent of the number of lower level lasers ($M$) or the number of HUs ($P$). Further simulations, not shown here, indicate that the exponent is also independent of the pumping current as long as it is sufficiently below threshold. For the lower layer lasers with current close to or above laser threshold, each laser can no longer be considered a "passive" element and large fluctuations in synchronization times are observed.

Since the critical coupling can now be identified with great precision, the nature of the phase transition to crowd synchrony, e.g. first or second order, can be addressed. To do so the coupling strength for an architecture with $P=3$ is slowly increased, by $\Delta\sigma = 3 \cdot 10^{-3} \sigma_c$ every $10\tau$, from $0.975\sigma_c$, through the critical point, to $\sigma = 1.0275\sigma_c$. Once the system is fully synchronized, the coupling is slowly decreased back to its initial value (Fig. 3b). A hysteresis loop is formed, as expected for a first order phase transition [14], whose width is ~1-2% of the normalized sigma (the width depends on the rate of modification of the coupling strength).

The hysteresis loop is expected to be robust to a very slow modification rate of the coupling strength which should inversely scale exponentially with the number of lower level lasers, $M$. To check this hypothesis we measure the life-time of the hysteresis loop as a function of $M$. The network is first synchronized at $\sigma \sim 1.01\sigma_c$ and then the coupling strength is abruptly decreased to be within the hysteresis loop, $\sigma=0.995\sigma_c$. We next measure the decay of synchronization as a function of time, Fig. 3c. Results indicate a good collapse of the data for all $M$ in the range [20,100] and the estimated decay exponent is almost a constant independent of $M$ as shown in Fig. 3d. This lack of dependence of the decorrelation time on the system size indicates an unusual *discontinuous dynamical transition* which might be in the spirit of dynamical first order phase transition measured for spin-glass systems with the lack of latent heat [15, 16].

We find that crowd synchrony can also emerge in a diluted system where the lower layer lasers are influenced by different subsets of HUs. Returning to the expanded bridge analogy, one can now imagine pedestrians walking on a random subset of the bridges. Denoting the average connectivity ratio (number of connections/$M \cdot P$) by $R$, and empirically simulating systems with various ($M$, $P$, $R$) yields that the critical coupling, $\sigma_c$, to crowd synchrony occurs at

$$\sigma_c \cong \frac{A}{M^{0.463} \cdot P^{0.492} \cdot R^{0.993}} \sim \frac{A}{(M \cdot P)^{0.5} R} \quad (4)$$

where A is a constant and $M \cdot P$ is the maximal number of connections ($R=1$). For given parameters ($M$, $P$, $R$), $\sigma_c$ was obtained using the same methods as shown in Fig. 3a. Figure 4a-b, with $P=4$, clearly indicates that $\sigma_c \sim M^{-0.5}$, both for a fully connected system, $R=1$, and for various dilutions, $R<1$. The dependence of $\sigma_c$ on $R$ and $P$ is estimated from the same figure where $M$ is fixed. In the partially connected systems, the connections can be chosen randomly or systematically under the constraint that all hidden units have nearly the same number of connections. For random connections, $\sigma_c$ is consistently larger than in the systematic case (larger constant A in eq. (4)), and there are greater fluctuations between simulations with identical parameters, however, the scaling is still valid.

As the system approaches synchronization, feedback from other lasers is the dominant factor in eqs. (1,2). Neglecting other factors, we note that:

$$|\dot{E}_i| \sim \sum_{k=1}^{P} \sigma_{ik} E_k(t-\tau) \sim \sigma R P \overline{E_k} \quad (5)$$

$$|\dot{E}_k| \sim \sum_{i=1}^{M} \sigma_{ik} E_i(t-\tau) \sim \sigma R M \overline{E_i} \quad (6),$$

where $\overline{E_k}/\overline{E_i}$ is the average of field over all connected P/M lasers. Substituting eq. (6) in eq. (5) one finds a self-consistent equation for $|\ddot{E}_i|$

$$|\ddot{E}_i| = \sigma R P \dot{\overline{E_k}} = \sigma^2 R^2 P N \overline{E_i}, \quad (7)$$

and $|\overline{E}_i(t)| = A e^{at}$, where

$$a = \sigma R \sqrt{PN} \quad (8)$$

We note that there is a critical coupling strength for which the feedback becomes the dominant factor in equations (1,2). Eq. (8) indicates that the critical coupling has to scale with $1/R(PN)^{0.5}$ in accordance with Eq. (4).

The question of whether a transition to crowd synchrony could occur in the case of detuning between the hubs in a diluted system is examined using the limited architecture of two hidden units with frequency detuning, $w_2=a$, as in Fig. 1b. Here, $L/M=0.2$ fraction of the lower layer lasers are connected to both hidden units, with a weaker coupling strength $\sigma_2$. For small frequency detuning between the hidden units ($a=0.5\pi$ rad/ns) Fig. 5 indicates that the correlation between the two groups increases with $\sigma_2$, until all the lasers are synchronized. Increasing the frequency detuning between HUs two phenomena are observed. First, the average correlation diminishes, secondly with a weak $\sigma_2$ it appears that the correlation remains close to zero, until a threshold is passed, e.g. for 50% detuning, Fig. 5 indicates that correlation remains close to zero up to $\sigma_2 \sim 0.1\sigma_1$. These limited results indicate that in the presence of variation among the features of the hubs and breaking symmetry among the lower layer elements, crowd synchrony appears for strong enough couplings, however, the first order phase transition disappears and might now be of the second order type.

In conclusion, we have shown that a two layer system of delay-coupled oscillators can synchronize when the number of elements or strength of coupling is large enough. This is also shown in the case of a partially connected system, where the elements are indirectly connected through several mediators. The necessary coupling strength acts as a power law, depending on the number of elements, hubs and connectivity ratio. Increasing the detuning between hubs causes the correlation of the entire system to decrease and the phase transition is continuous and not of the first order type.

The authors thank for valuable comments by Rajarshi Roy and Wolfgang Kinzel.

# References

[1]    T. Danino, O. Mondragon-Palomino, L. Tsimring, and J. Hasty, Nature **463**, 326 (2010).

[2]    J. Garcia-Ojalvo, M. B. Elowitz, and S. H. Strogatz, Proc. Nat. Acad. Sci. **101**, 10955 (2004).

[3]    K. Wiesenfeld, C. Bracikowski, G. James, and R. Roy, Phys. Rev. Lett. **65**, 1749 (1990).

[4]    S. H. Strogatz, *Sync: The emerging science of spontaneous order* (Hyperion, 2003).

[5] P. Dallard *et al.*, Structural Engineer **79**, 17 (2001).

[6] J. Zamora-Munt, C. Masoller, J. Garcia-Ojalvo, and R. Roy, Phys. Rev. Lett. **105**, 264101 (2010).

[7] S. De Monte, F. d'Ovidio, S. Danø, and P. G. Sørensen, PNAS **104**, 18377 (2007).

[8] A. F. Taylor, M. R. Tinsley, F. Wang, Z. Huang, and K. Showalter, Science's STKE **323**, 614 (2009).

[9] A. Engel, and C. Broeck, *Statistical mechanics of learning* (Cambridge Univ Press, 2001).

[10] M. Opper, and W. Kinzel, in *Models of neural networks III*, edited by E. Domany, J. L. van Hemmen, and K. Schulten (Springer-Verlag, Berlin, 1996), pp. 151.

[11] F. Rosenblatt, Psychological review **65**, 386 (1958).

[12] M. Opper, EPL **8**, 389 (1989).

[13] A. Priel, M. Blatt, T. Grossmann, E. Domany, and I. Kanter, Phys. Rev. E **50**, 577 (1994).

[14] G. Bertotti, *Hysteresis in magnetism: for physicists, materials scientists, and engineers* (Academic Pr, 1998).

[15] T. R. Kirkpatrick, and D. Thirumalai, Phys. Rev. Lett. **58**, 2091 (1987).

[16] D. J. Gross, I. Kanter, and H. Sompolinsky, Phys. Rev. Lett. **55**, 304 (1985).

# FIGURES

**Fig. 1(color online).** (a) Schematic of *M* non-identical elements interacting with each other via *P* hidden units (HUs), where $\sigma_{ik}$ stands for the coupling strength between element *i* and the *kth* HU. A dilution consists of setting a fraction of the couplings to vanishing values. (b) Schematic of an architecture with 2M elements and two non-identical HUs, with frequency detuning, $w_i$ between them, where only 2L out of the 2M elements have couplings to both HUs. Singly connected elements have coupling strengths $\sigma_1$ while elements coupled to two HUs have coupling strengths $\sigma_2$.

**Fig. 2 (color online).** Color chart of intensity correlation among all pair of lasers in the lower layer, $\rho(i,j)$, for the architecture of Fig. 1a with *M*=20 and *P*=3 for three different coupling strengths. The lower layer lasers are sorted by increasing frequency. (a) $\sigma$=24.55 where all pair correlations are below the threshold (below criticality). (b) $\sigma$=24.67, correlation begins to form. (c) $\sigma$=24.78, all pairs of lasers are correlated. The transition to crowd synchrony is identified at $\sigma_c$=24.6398 as explained in the text.

**Fig.3 (color online).** (a) A power law behavior of synchronization time as a function of the deviation from $\sigma_c$. (b) Average correlation among all pairs of lower level lasers as a function of the normalized sigma ($\sigma/\sigma_c$) shows a hysteresis loop. (c) Decay of the correlation as a function of time, where $\sigma$ is abruptly changed from a synchronized state, $\sigma>\sigma_c$, to $\sigma$=0.995$\sigma_c$ (see text for detail). Results indicate data collapse and are obtained for *P*=3 and *M* in the range [20,100]. (d) Correlation decay exponent as a function of M obtained from the data of panel c.

**Fig.4 (color online).** (a) Critical coupling as a function of number of lower layer lasers, for different number of hidden units with the lack of dilution. (b) Critical coupling as a function of number of lower layer lasers, for different connectivity ratio, *R*, and with *P*=4. The dashed lines are given by the middle expression of eq. 4 with *A*~169.6.

---

**Fig. 5 (color online).** Correlation among all lower layer lasers in Fig. 1b as a function of $\sigma_2$ $\sigma_2$, for different frequency detuning between the two HU lasers. As detuning is increased, the average correlation weakens. M = 40/80, L = 0.2M, *$\sigma_2$=1.3$\sigma_c$*.

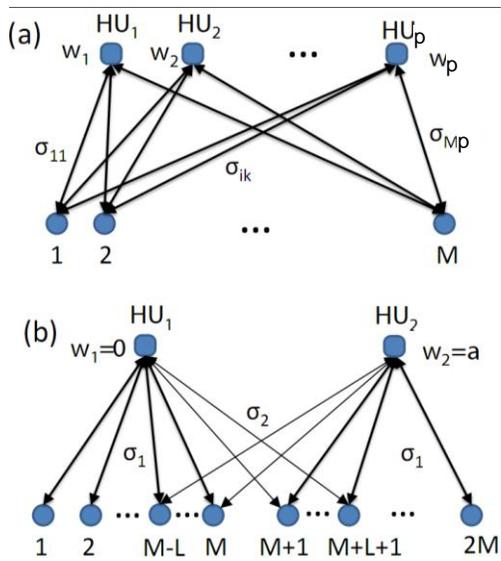
Figure 1

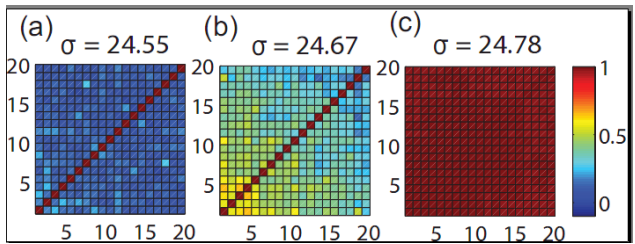
Figure 2

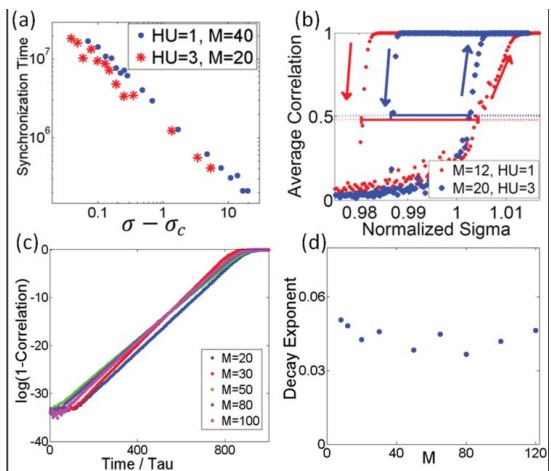
Figure 3

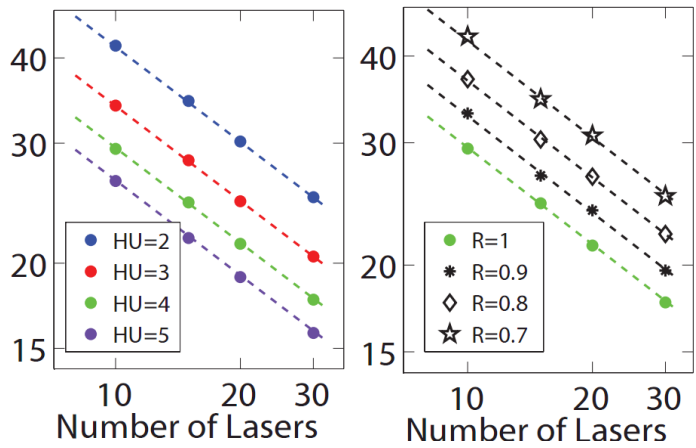

Figure 4

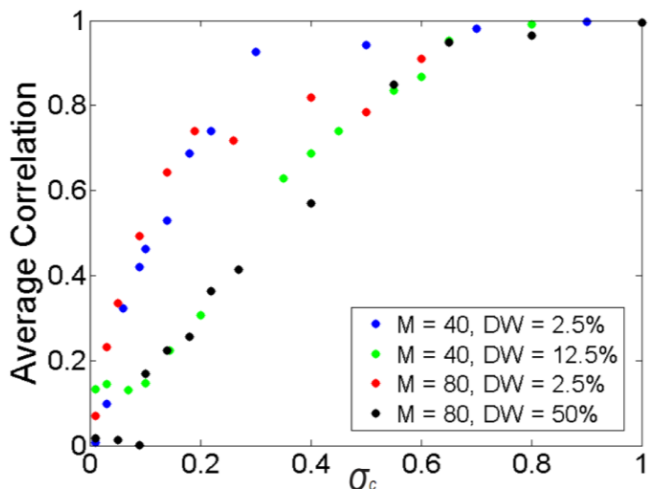

Figure 5